\documentclass[10pt]{iopart}

\usepackage{graphicx}
\usepackage{psfrag}
\usepackage{color}
\usepackage{amsbsy}
\usepackage{amssymb}
\usepackage{iopams}  

\def\a{\alpha}
\def\b{\beta}
\def\g{\gamma}
\def\d{\delta}

\def\e{\epsilon}

\def\s{\sigma}

\def\S{{\bf S}}

\begin{document}

\title[Proton conduction]
{An interacting spin flip model for one-dimensional proton conduction}

\author{Tom Chou}

\address{Dept. of Mathematics, Stanford University, Stanford, CA 94305-2125
\footnote{Present Address: Dept. of Biomathematics $\&$ Dept. of Physics, 
UCLA, Los Angeles, CA 90095}}

\begin{abstract}

A discrete asymmetric exclusion process (ASEP) is developed to
model proton conduction along one-dimensional water wires.  Each
lattice site represents a water molecule that can  be in only one
of three states; protonated, left-pointing, and right-pointing.
Only a right(left)-pointing water can accept a proton from its
left(right).  Results of asymptotic mean field analysis and
Monte-Carlo simulations for the three-species, open boundary
exclusion model are presented and compared.  The mean field
results for the steady-state proton current suggest a number of
regimes analogous to the low and maximal current phases found in
the single species ASEP [B.  Derrida,  Physics Reports, {\bf 301},
65-83, (1998)].  We find that the mean field results are accurate
(compared with lattice Monte-Carlo simulations) only in the
certain regimes.  Refinements and extensions including more
elaborate forces and pore defects are also discussed.

\end{abstract}


\submitto{\JPA}


\section{Introduction}

One-dimensional nonequilibrium processes have been used as models for many
systems including interface growth \cite{KRUG,KANDEL}, traffic flow
\cite{TRAFFIC,TRAFFIC2}, pore transport \cite{PORE1,PORE2,PORE3}, and
quantum spin chains \cite{SANDOW}. The simple asymmetric exclusion process
(ASEP) has also been used to model mRNA translation \cite{RNA1,RNA2} and
recent modifications have been applied in order to describe molecular motors
and parallel pathways \cite{MOTORREV,TUBE,KOLO}.

Building upon the exact and asymptotic results of Derrida, Evans, Rittenberg,
Sandow, and others
\cite{SANDOW,DER92,DER93,DER97,DER98,RITTENBERG1}, the quantitative
behavior of other more complex one-dimensional systems can be investigated. 
Asymmetric exclusion processes of multiple species have also been modelled,
particularly for periodic boundary conditions \cite{RAJEWSKY99,IRAN99}, or in
the context of spontaneous symmetry breaking for open systems
\cite{EVANS95}.  In the proton conduction problem, only one species (protons)
are transported while the other classes of particles correspond to different 
orientation states of the
relatively fixed water molecules.

In this paper, we present a restricted three-species exclusion process that models
proton conduction through a ``water wire.'' Our model is motivated by experiments on
proton conduction through stable or transient water filaments suggesting conduction
occurs via proton exchange between properly aligned water molecules
\cite{EISENMAN80,DEAMER2}.  This ``Grotthuss \cite{GROTTHUSS}'' mechanism of
proton conduction along a relatively static water chain has been proposed to give rise
to the fast proton transfer rates relative to the permeation of distinguishable ions
across structurally similar ion channels.

\section{Lattice Model}

Figure 1a is a schematic of model in which each ``site'' along the pore is
occupied by an oxygen atom.  This oxygen is either part of neutral water,
H$_{2}$O, or hydronium H$_{3}$O$^{+}$.  We assume hydronium ions are
spherically symmetric, in the sense that their three protons are either in a planar,
left-right symmetric hybridization, or are rapidly smeared over the molecule.
These singly protonated species are denoted ``$0$'' type particles.  In contrast,
the neutral waters have ``permanent'' dipole moments or electron lone-pair
orientations that can nonetheless fluctuate. For simplicity, all water dipoles
(hydrogens) that point to right are classified as ``$+$'' particles, while those
pointing left are ``$-$'' particles.  Thus, each site can exist in only one of three
states: $0, +$, or $-$, corresponding to protonated, right, or left states,
respectively.  This simple choice of for labeling the occupancy configurations,
$\s_{i} = \{-1, 0, +1\}$, allows for fast integer computation in simulations.

\begin{figure}
\begin{center}
\includegraphics[height=2.4in]{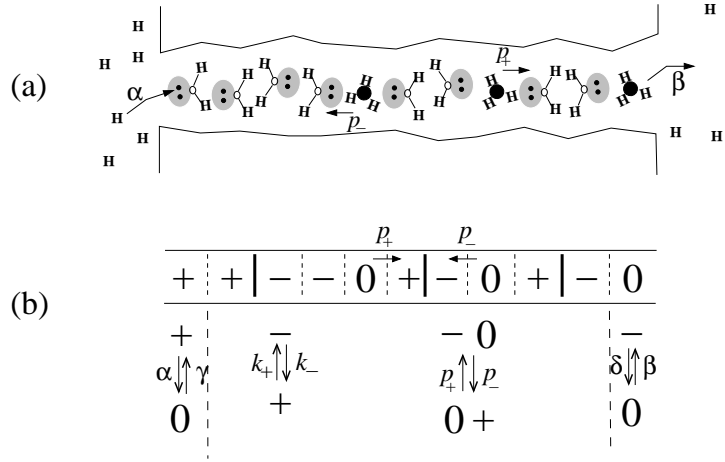}
\caption{(a) Schematic of an $N=11$ three-species exclusion model 
that captures the steps in a Grotthuss mechanism of proton transport
along a water wire. For typical gramicidin pores that span 
lipid membranes $N\sim 20-22$. The transition rates are labeled in (b).
Water dipole kinks are denoted by thick lines.}
\end{center}
\label{fig1}
\end{figure}

The transition rules are constrained by the orientation of the waters at each site
and are defined in Fig. 1b.  A proton will enter the first site ($i=1$) from the left
reservoir and protonate the first water molecule with rate $\a$ only if the
hydrogens of the first water is pointing to the right (such that its lone-pair
electrons are left-pointing, ready to accept proton from the left reservoir). 
Similarly, if a proton exits from the first site into the left reservoir (with rate
$\gamma$), it leaves the remaining hydrogens right-pointing. The analogous
rules hold for the last site $i=N$ with exit and entrance rates $\beta$ and
$\delta$ respectively. The entrance rates $\alpha$ and $\delta$ are functions to
the proton concentration in the respective reservoirs as well as local electric
potential gradients near the pore mouths.  In the pore interiors, a proton at site
$i$ can hop to the right(left) with rate $p_{+}(p_{-})$ only if the adjacent particle
is a right(left)-pointing, unprotonated water molecule. If such a transition is
made, the remaining water molecule at site $i$ will be left(right) pointing. 
Physically, as a proton moves to the right, it leaves a decaying trail of $-$
particles to its left.  A left moving proton leaves a trail of $+$ particles to its
right.  These trails of $-$ or $+$ particles are unable to accept another proton
from the same direction.  Clearly, protons can successively follow each other
only if the waters can reorient without being protonated. The water reorientation
rates are denoted $k_{\pm}$ (cf. Fig. 1b).  The rate-limiting steps in steady-state
proton transfer across biological water channels is thought to be associated with
water flipping.  

This water wire conduction model resembles the two species ASEPs treated by Evans
{\it et al.} \cite{EVANS95} if one identifies protons with vacancies, and allows the $+$
and $-$ clkass particles to interconvert with rates $k_{\pm}$. Moreover, $+$ particles
here can only move left, and once they do, get converted to $-$ type particles.
Similarly, a $-$ particle that moves to the right becomes a $+$ particle. The allowed
moves are $- 0 \leftrightarrow 0 +$.  Although the master equation can be formally
defined in terms of the probability $P_{N}(\s_{1}, \s_{2}, \ldots \s_{N-1},\s_{N})$ for a
given configuration $\{ \s_{i} \}$, I have failed to discover an analytical result for the
steady state current proton ($\s = 0$) current using matrix product methods
\cite{SANDOW,DER93,DER98,RITTENBERG1}. Since it is not always possible to find
a finite dimensional matrix product representation, particularly for open boundaries,
many practical applications, such as multipecies pore transport \cite{PORE3}, must be
treated using mean field approximations or numerical simulations. For the single
species ASEP, the mean field result for the steady-state current is exact in the
$N\rightarrow \infty$ limit, although the site occupancies do not agree with those from
the exact solution. Therefore, a mean field theory in the present problem may shed
light on the steady-states. We will obtain mean field analytic expressions
that can be compared with numerical simulations. 

\section{Mean field approximation}

For one-dimensional nearest lattice site hopping models, the mean-field assumption
states that the probability that the system is in a configuration with occupancy $\s_{i}$
and $\s_{i+1}$ factors into a product of single particle probabilities:

\begin{equation}
\fl \sum_{\tau_{i}=0,\pm 1}\!\!
P_{N}(\tau_{1},\ldots,\s_{i},\s_{i+1},\ldots,\tau_{N})  \equiv P_{2}(\s_{i},\s_{i+1}) 
\approx P(\s_{i})P(\s_{i+1}),
\end{equation}

\noindent where $P(\s_{i}) \equiv \sum_{\s_{i+1}=0,\pm 1}\sum_{\{\tau_{i}\}=0,\pm 1}
P_{N}(\tau_{1},\ldots,\s_{i},\s_{i+1},\ldots,\tau_{N})$.
The master equations thus become

\begin{eqnarray} 
\fl \dot{P}(\s_{i}=0) = -p_{+} P(\s_{i}=0)P(\s_{i+1}=+1)-p_{-}P(\s_{i}=0)P(\s_{i-1}=-1) \nonumber \\
\lo \quad +p_{+}P(\s_{i-1}=0)P(\s_{i}=+1)+p_{-}P(\s_{i+1}=0)P(\s_{i}=-1)\\[13pt]
\fl \dot{P}(\s_{i}=+1)=  -k_{+}P(\s_{i}=+1)+k_{-}P(\s_{i}=-1)\nonumber\\
\lo  \quad -p_{+}P(\s_{i}=+1)P(\s_{i-1}=0)+p_{-}P(\s_{i}=0)P(\s_{i-1}=-1)\\[13pt]
\fl \dot{P}(\s_{i}=-1) =  k_{+}P(\s_{i}=+1)-k_{-}P(\s_{i}=-1)\nonumber \\
\lo  \quad -p_{-}P(\s_{i+1}=0)P(\s_{i}=-1)+p_{+}P(\s_{i}=0)P(\s_{i+1}=+1),
\end{eqnarray}

\noindent where $P(\s_{j}=\s)$ are the individual probabilities that site $j$ is occupied
by a particle in state $\s$.  Conservation at each site requires
$P(\s_{i}=-1)+P(\s_{i}=0)+P(\s_{i}=+1) = 1$.  In the present treatment, we will assume
that $p_{\pm}, k_{\pm}$ are independent of neighboring configurations $\s$, {\it i.e.}
that the transition rates are local and do not depend interaction with the states of
neighboring sites. Within the mean field approximation, the instantaneous current of
protons between sites $i$ and $i+1$ is

\begin{equation}
\fl \quad J(i,i+1) = p_{+}P(\s_{i}=0)P(\s_{i+1}=+1)-p_{-}P(\s_{i+1}=0)P(\s_{i}=-1)
\end{equation}

\noindent while the currents at the left and right boundaries are

\begin{equation}
\begin{array}{l}
J(L,1) = \a P(\s_{1}=+1) -\g P(\s_{1}=0)  \\[13pt]
J(N,R) = \b P(\s_{N}=0) -\d P(\s_{N}=-1).
\label{BC0}
\end{array}
\end{equation}

Upon defining steady-state particle occupancies over their probabilities, $\langle
P(\s_{i}=0) \rangle \equiv r_{i}$, $\langle P(\s_{i}=+1) \rangle = s_{i}$, 
and $\langle P(\s_{i}=-1)\rangle \equiv t_{i}$. The steady-state currents are thus

\begin{equation}
\langle J(i,i+1)\rangle = p_{+}r_{i} s_{i+1} -p_{-}r_{i+1}t_{i}
\label{JIAVE}
\end{equation}

\noindent and

\begin{equation}
\begin{array}{l}
\langle J(L,1)\rangle = \a s_{1}-\g r_{1} = (\a+\g)s_{1}+\g t_{1}-\g \\[13pt]
\langle J(N,R) \rangle= \b r_{N}-\d t_{N} = \b-\b s_{N} -(\b+\d)t_{N}.
\label{BCJ}
\end{array}
\end{equation}


\noindent In steady-state, all currents have to be identical,
$J = \langle J(i,i+1)\rangle = \langle J(L,1)\rangle = \langle J(N,R)\rangle$. 
Thus, the constant, internal steady-state current (\ref{JIAVE}) reads 

\begin{equation}
J \equiv \langle J(i,i+1)\rangle = p_{+} r_{i} s_{i+1}-p_{-} r_{i+1} t_{i},
\label{JJ}
\end{equation}

\noindent while the equation for 
$\langle \dot{P}(\s_{i}=+1)\rangle = 0$ reads

\begin{equation}
0=k_{+}s_{i}-k_{-}t_{i}-p_{-}r_{i+1}t_{i}+p_{+}r_{i}s_{i+1} = k_{+}s_{i}-k_{-}t_{i}+J.
\label{TAU=0}
\end{equation}

\noindent From (\ref{TAU=0}), 
we see that $k_{+}s_{i}-k_{-}t_{i} = -J$ is constant along the chain.
Upon imposing particle conservation, $r_{i}+s_{i}+t_{i}=1$,

\begin{equation}
t_{i} = Ks_{i} + {J\over k_{-}} = {K(1-r_{i})\over K+1} + {J \over k_{+}+k_{-}},
\quad K\equiv {k_{+}\over k_{-}}.
\label{TI}
\end{equation}

\noindent The mean end-site occupations can be found by substituting 
(\ref{TI}) into (\ref{BCJ}), yielding

\begin{equation}
r_{1} = {\a \over \a+\g (K+1)}-{K+1+\a/k_{-} \over \a +\g (K+1)}J,
\label{R1}
\end{equation}

\begin{equation}
r_{N}= {K\d \over \b(K+1)+K\d} + {K+1+\d/k_{-} \over \b(K+1)+K\d}J.
\label{RN}
\end{equation}


For the internal sites, substitution of (\ref{TI}) into (\ref {JJ})
yields

\begin{equation}
0 = A_{4}r_{i}r_{i+1} + A_{3}r_{i+1}-A_{2}r_{i}-A_{1},
\label{A1234}
\end{equation}

\noindent where 

\begin{equation}
\fl A_{1} = J,\,  A_{2} =  \displaystyle p_{+}\left[{k_{-}-J\over k_{+}+k_{-}}\right]\!,\,
A_{3} = \displaystyle p_{-}\left[{k_{+}+J \over k_{+}+k_{-}}\right]\!,\,\,\, \mbox{and}\,\,\,
\displaystyle A_{4} =  {k_{-}p_{+}-k_{+}p_{-} \over k_{+}+k_{-}}. 
\label{A}
\end{equation}


\noindent We can now write (\ref{A1234}) as a recursion relation of the form

\begin{equation}
r_{i+1} = {A_{2}r_{i}-A_{1} \over A_{4}r_{i}+A_{3}}.
\label{Si+1}
\end{equation}

\noindent Upon redefining 

\begin{equation}
r_{i} \equiv {A_{2}+A_{3}\over A_{4}}\xi_{i} - {A_{3}\over A_{4}},
\label{RXI}
\end{equation}

\noindent we find the recursion relation for $\xi$:

\begin{equation}
\xi_{i+1} = 1 - {c(J)\over \xi_{i}},
\quad c(J) \equiv {A_{2}A_{3}+A_{1}A_{4} \over (A_{2}+A_{3})^{2}}.
\label{XIRECURSION}
\end{equation}

\noindent Except for the more complicated dependence of $c(J)$ on $J$, the
recursion relation (\ref{XIRECURSION}) is identical to that used in analyzing the
mean-field properties of the totally and partially asymmetric exclusion
process for a single species \cite{DER92,RITTENBERG1}.  The solution to the
difference equation (\ref{XIRECURSION}) is \cite{DER92}

\begin{equation}
\xi_{k} = {(\xi_{+}^{k}-\xi_{-}^{k})\xi_{1} - \xi_{+}\xi_{-}(\xi_{+}^{k-1}
-\xi_{-}^{k-1}) \over (\xi_{+}^{k-1}-\xi_{-}^{k-1})
\xi_{1}-\xi_{+}\xi_{-}(\xi_{+}^{k-2}-\xi_{-}^{k-2})},
\end{equation}

\noindent where

\begin{equation}
\xi_{\pm} = {1\over 2}\left[1\pm \sqrt{1-4c(J)}\right].
\end{equation}

\noindent Similarly, the boundary occupations $\xi_{1},\xi_{N}$ can be readily found 
by substituting (\ref{R1}) and (\ref{RN}) into (\ref{RXI}) and solving for $\xi$.

\subsection{Symmetric limit}

First consider the totally symmetric case where $\vert k_{-}p_{+}-k_{+}p_{-}\vert < o(1/N)$ and the
``unbiased'' current vanishes as $1/N$.  One important realisation of this condition is
$k_{+}/k_{-}=K=1$ and $p_{+}=p_{-}$, corresponding to waters without left-right preference along the
wire, and to the absence of pondermotive forces on the wire protons.  We name this limit the symmetric
limit because the net bias on internal protons vanish \footnote[1]{In the proton conduction literature,
the term ``symmetric'' is often used to denote symmetric solutions, or equal proton concentrations in
the two reservoirs. The current in this case is driven by an externally imposed transmembrane
potential}.  The proton current is driven entirely by a proton number density difference between the
two reservoirs, {\it i.e.} an asymmetry in injection and extraction rates at the {\it boundaries}. 
Upon iterating Eqn (\ref{Si+1}) with $A_{4} \propto k_{-}p_{+}-k_{+}p_{-} = 0$,

\begin{equation}
r_{N} = \left({A_{2}\over A_{3}}\right)^{N-1}r_{1} - \left({A_{1}\over A_{2}}\right){
1-(A_{2}/A_{3})^{N-1}\over 1-(A_{2}/A_{3})}.
\end{equation}

\noindent From the boundary conditions at $i=1,N$, 
we obtain an implicit equation for $J$:

\begin{equation}
\begin{array}{l}
\fl \displaystyle \left({A_{2}\over A_{3}}\right)^{N-1}
\left[{\a \over \a+\g (K+1)}-{K+1+\a/k_{-} \over \a +\g (K+1)}J\right]
- {A_{1}\over A_{3}}
{1 - (A_{2}/A_{3})^{N-1}\over 1-(A_{2}/A_{3})} = \\[13pt]
\fl \displaystyle \hspace{5.5cm} 
{K\d \over \b(K+1)+K\d} + {K+1+\d/k_{-} \over \b(K+1)+K\d}J.
\label{IMPLICIT}
\end{array}
\end{equation}

\noindent  where the $A_{i}$ are functions of $J$ given by Eqns.
(\ref{A}). A closed analytic  expression can be found as an asymptotic expansion about
large $N$ using the {\it ansatz}

\begin{equation}
J = {a_{1} \over N} + {a_{2} \over N^{2}} + O\left({1\over N^{3}}\right),
\end{equation}

\noindent where the $a_{i}$ are $N$-independent 
coefficients that are functions of the kinetic rates.
To lowest order,

\begin{equation}
\left({A_{2}\over A_{3}}\right)^{N-1}
\sim \left(1-a_{1}{K+1\over N k_{+}}\right)^{N-1} \sim
\exp\left(-{a_{1} (K+1)\over k_{+}}\right).
\end{equation}

\noindent At $O(1/N^{0})$, (\ref{IMPLICIT}) determines 
$a_{1}$ and the steady-state current

\begin{equation}
\fl J = {k_{+}k_{-}\over N(k_{+}+k_{-})}
\ln \left[{\b(K+1)+K\d\over \g(K+1)+\a}{\g(k_{+}+k_{-})+\a(p_{-}+k_{-})\over
\b(k_{+}+k_{-})+K\d(p_{-}+k_{-})}\right] + O(N^{-2}).
\label{J2}
\end{equation}

\noindent When $\b=\g$, the current (\ref{J2}) in the limit of small 
proton concentration differences can be expanded in powers of 
$k_{-}\a - k_{+}\d$,

\begin{equation}
\fl J \sim {\b p_{+}(k_{-}\a - k_{+}\d) \over 
N \left[\b(K+1)+K\d\right](\b+\d) (k_{+}+k_{-})} + O\left((k_{-}\a-k_{+}\d)^{2}\right)
+O(1/N^{2}),
\label{J2b}
\end{equation}

\noindent reflecting a current driven by the effective difference 
in boundary injection rates.  Finally, in the {\it large} $\a$ and $\d = 0$ limit, 

\begin{equation}
\fl J \sim {k_{+}k_{-} \over (k_{+}+k_{-})N} \log \left(1+{p_{-}\over k_{+}}\right) 
- {\gamma k_{+}k_{-}p_{-} \over 
\a N (k_{+}+k_{-})(k_{+}+p_{-})} + O(\a^{-2}N^{-1}).
\end{equation}

\subsection{Asymmetric limits}

For driven systems, where $\vert k_{-}p_{+}-k_{+}p_{-}\vert > O(1/N)$, a finite current 
persists in the $N\rightarrow \infty$ limit. Mean field
currents are found from the fixed points of the recursion relation (\ref{XIRECURSION}).
The average particle occupations at each site $i$ can be then be reconstructed from the
solution to $\xi_{i}$.  In analogy to the single-species partially asymmetric process,
the condition $c(J_{M}) = 1/4$ leads to a single fixed point that determines $J_{M}$, via a
quadratic equation which is solved by


\begin{equation}
\begin{array}{l}
\fl \displaystyle  J_{M}^{\pm} = {2(p_{+}k_{-}-p_{-}k_{+}) \over (p_{+}+p_{-})^{2}}
\left[{(p_{-}+p_{+})\over 2}+k_{-}+k_{+}\pm\sqrt{k_{+}+
k_{-}}\sqrt{k_{-}+k_{+}+p_{+}+p_{-}}\right] \\[13pt]
\:\hspace{4cm} \mbox{when}\quad \xi_{1}(J_{M}^{\pm})\geq {1\over 2}, \, 
\xi_{N}(J_{M}^{\pm})\leq {1\over2}.
\label{JM}
\end{array}
\end{equation}

\noindent For the sake of simplicity, we assume $k_{-}p_{+}-k_{+}p_{-} > 0$ and furthermore
$\{\a, \b, \g, \d, k_{\pm}\}$ are such that $J > 0$. Clearly $J<k_{-}$ 
(since $J$ must be less than the smallest
rate in the entire process) and $A_{4}>0$. In the
limit $p_{+}/(k_{+}+k_{-}), p_{-}/(k_{+}+k_{-})\rightarrow 0$, only $J_{M}^{-}$ provides a
physical solution,


\begin{equation}
J_{M}^{-} \sim {p_{+}k_{-}-p_{-}k_{+}\over k_{+}+k_{-}}
\left[{1\over 4}-{p_{+}+p_{-}\over 8(k_{+}+k_{-})}+ 
O\left(\left({p_{+}+p_{-} \over k_{+}+k_{-}}\right)^{2}\right)\right].
\end{equation}

\noindent For a purely asymmetric process $p_{-}=0$ and 
the maximal current $J_{M}^{-}$ in this model approaches the analogous maximal current 
expression of the single species ASEP,

\begin{equation}
J_{M}^{-}(p_{-}=0) \sim {p_{+} \over 4(K+1)} + O\left({p_{+}\over k_{-}}\right),
\end{equation}

\noindent except for modification by the factor $(K+1)^{-1}$ representing the
fraction of time the site to the immediate right of a proton is in the $+$ state.

Another possibility is $c(J<J_{M}^{\pm}) < 1/4$ which leads to two fixed points, one stable at
$\xi_{-}$ and one unstable at $\xi_{+}$. Following \cite{DER92} and \cite{RITTENBERG1},
we set in one case the left occupation $\xi_{1} = \xi_{-}$ to the stable value which
self-consistently determines the current via $c(J) = \xi_{1}(1-\xi_{1})$, while
equation (\ref{R1}) determines $\xi_{1}(J)$. Thus $c(J_{L}) \equiv
\xi_{1}(J_{L})(1-\xi_{1}(J_{L}))$ is solved by 

\begin{equation}
\fl J_{L}^{\pm} = {G(\a,\g,k_{\pm},p_{\pm})\over 2 F(\a,\g,k_{\pm},p_{\pm})}
\pm {\sqrt{G^{2}(\a,\g,k_{\pm},p_{\pm})-4
H(\a,\g,k_{\pm},p_{\pm})F(\a,\g,k_{\pm},p_{\pm})}\over 2 F(\a,\g,k_{\pm},p_{\pm})},
\label{JL}
\end{equation}

\noindent where 

\begin{equation}
\begin{array}{l}
\fl F(x,y,k_{\pm},p_{\pm}) \equiv (k_{+}+k_{-})(x+k_{+}+k_{-})
\big[y(p_{+}+p_{-})+x p_{-}-U\big]U \\[13pt]
\fl G(x,y,k_{\pm},p_{\pm}) \equiv (k_{+}+k_{-})U \bigg[x^{2}k_{-}
(k_{-}+p_{-})-(x k_{-} -y (k_{+}+k_{-})) U + \\
\fl \hspace{2.7cm} xy (2k_{+}k_{-}+2k_{+}p_{+}+k_{-}p_{-}
-k_{+}p_{-}+2k_{-}^{2})+y^{2}(k_{+}+k_{-})^{2}\bigg] \\[13pt]
\fl H(x,y,k_{\pm},p_{\pm}) \equiv xy k_{-}(k_{-}+p_{+})U^{2}(k_{\pm},p_{\pm}) \\[13pt]
\fl U(k_{\pm},p_{\pm}) \equiv k_{-}p_{+}-k_{+}p_{-}.
\end{array}
\label{FGH}
\end{equation}

\noindent The injection rate-limited current (\ref{JL}) depends only upon $\a, \g, k,
p_{\pm}$.  Upon iterating, we find the regime of validity for $J \sim J_{L}$ is
determined by $\xi_{N} < \xi_{+}$. Thus, the currents $J_{L}^{\pm}$ can arise only when
$\xi_{1}(J_{L}^{\pm}) < 1/2$ and $\xi_{N}(J_{L}^{\pm}) < 1-\xi_{1}(J_{L}^{\pm})$ and
correspond to a low proton density ($r_{i}$) phase.


\begin{figure}
\begin{center}
\includegraphics[width=5.5in]{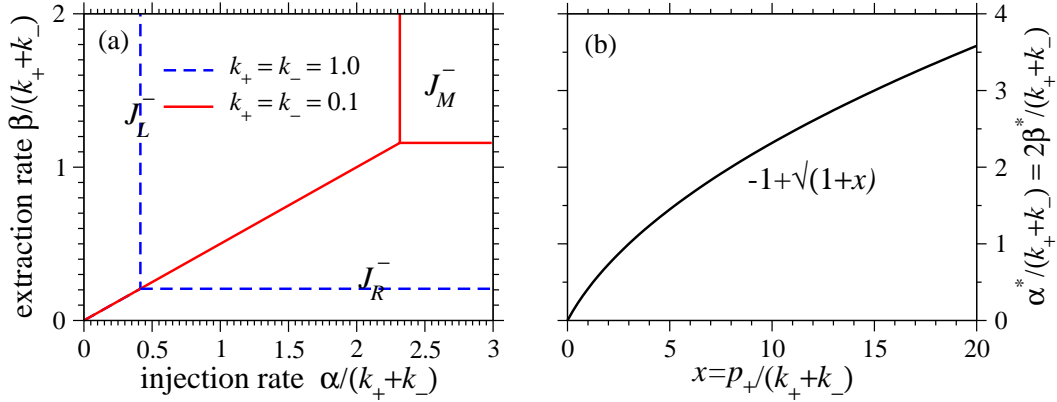}
\caption{(a). Mean field phase diagram in the totally asymmetric limit
$\delta = \gamma = p_{-} = 0$. Three current regimes for 
$p_{+}/(k_{+}+k_{-}) = 10$ are 
delineated by thick solid curves at  values of $\a^{*}/(k_{+}+k_{-}) = 
2\b^{*}/(k_{+}+k_{-}) = \sqrt{1+x}-1$. Thin dashed lines define the 
mean field phase diagram for $k_{+}=k_{-}=1.0$. (b). The dependence of 
$\a^{*}/(k_{+}+k_{-})$ on $x=p_{+}/(k_{+}+k_{-})$.}
\end{center}
\label{PHASE}
\end{figure}

Another limit of steady-state currents corresponds to the high density phase 
where $\xi_{N} = \xi_{+}$, which is found from $\xi_{N}(J_{R})(1-\xi_{N}(J_{R}))
\equiv c(J_{R}<J_{M})$ with 

\begin{equation}
\fl \displaystyle J_{R}^{\pm} = -{G(\d,\b,k_{\mp},p_{\mp})\over 
2 F(\d,\b,k_{\mp},p_{\mp})} \pm {\sqrt{G^{2}(\d,\b,k_{\mp},p_{\mp})-4
H(\d,\b,k_{\mp},p_{\mp})F(\d,\b,k_{\mp},p_{\mp})} \over 2 F(\d,\b,k_{\mp},p_{\mp})}.
\label{JR}
\end{equation}

\noindent Equation (\ref{JR}) is also 
easily deduced from the solutions $J_{L}^{\pm}$ and the symmetry properties of
$c(J), r_{1}$, and $r_{N}$ under the transformation $\{\a,\g,k_{\pm},p_{\pm}, J\} \rightarrow
\{\d,\b,k_{\mp},p_{\mp}, -J\}$.  The solutions $J_{R}^{\pm}$ hold only if
$\xi_{N}(J_{R}^{\pm})>1/2, \,\xi_{N}(J_{R}^{\pm})>1-\xi_{1}(J_{R}^{\pm})$ and correspond to a high
proton density phase. Although the currents represented by equations (\ref{JM}), (\ref{JL}),
(\ref{JR}), and their associated regimes of validity define a potentially rich
mean field current
phase diagram, significant simplification, where simpler expressions can be obtained, is
achieved by restricting ourselves to the totally asymmetric regime where $\g = \d = p_{-} = 0$.
In this case, $A_{3}=0, A_{4} = p_{+}/(K+1)$, and $\xi = k_{-}r/(k_{-}-J)$. Upon using $r_{1}$
and $r_{N}$ given by (\ref{R1}) and (\ref{RN}), the currents (\ref{JM}), (\ref{JL}), and
(\ref{JR}), and their regimes of validity become

\begin{equation}
\begin{array}{cl}
\fl \displaystyle J_{M}^{-} = k_{-}\left[1+{2(k_{+}+k_{-})\over p_{+}}
\left(1-\sqrt{1+{p_{+}\over k_{+}+k_{-}}}\right)\right]
&\displaystyle \quad \displaystyle \a \geq \a^{*},  \b \geq \b^{*}\\[13pt]
\fl \displaystyle J_{L}^{-} = {\a k_{-}(1-\a/p_{+})\over \a+k_{+}+k_{-}} & 
\displaystyle \quad \a \leq \a^{*}, \b\geq {\a \over (K+1)} \\[13pt]
\fl \displaystyle J_{R}^{-} = \b {k_{-}(1-\b/p_{+})-\b k_{+}/p_{+} \over \b+k_{-}} & 
\displaystyle \quad \b \leq \b^{*}, \a \geq \b(K+1),
\label{J0}
\end{array}
\end{equation}

\noindent where 

\begin{equation}
\begin{array}{l}
\displaystyle \a^{*} \equiv (K+1)\b^{*} \equiv 
(k_{+}+k_{-})\left(\sqrt{1+{p_{+}\over k_{+}+k_{-}}}-1\right).
\label{ABSTAR}
\end{array}
\end{equation}

\noindent The phase diagram defined by (\ref{J0}) and (\ref{ABSTAR}) is shown in
Fig. 2a for $x\equiv p_{+}/(k_{+}+k_{-}) = 1,10$. In the $k_{\pm}\rightarrow
\infty$ limit, the currents (\ref{J0}) reduce to modified
forms of the standard currents in the single particle ASEP \cite{DER97,DER98}.

\section{Monte-Carlo simulations}

We now compare  mean field results from both the symmetric and asymmetric
limits with their corresponding steady-state currents obtained from lattice
Monte Carlo (MC) simulations. The simulation was implemented by defining
$N+2$ sites corresponding to the interior pore sites, plus the two reservoirs. At
each time step, one of the $N+2$ sites was randomly chosen and a transition
attempted. Transitions were accepted according to the Metropolis algorithm. 
At the next time step, another site was randomly chosen.  The occupation at each
reservoir end site was set to $\s_{0}=\s_{N+1}=0$ so that protons were always
ready to be injected with probability corresponding to the rates $\a, \d$.  To
obtain low noise steady-state values for the occupations and currents,
instantaneous currents across all boundaries were averaged over
$O(10^{8}-10^{9})$ time steps.  

\begin{figure}
\begin{center}
\includegraphics[width=5.5in]{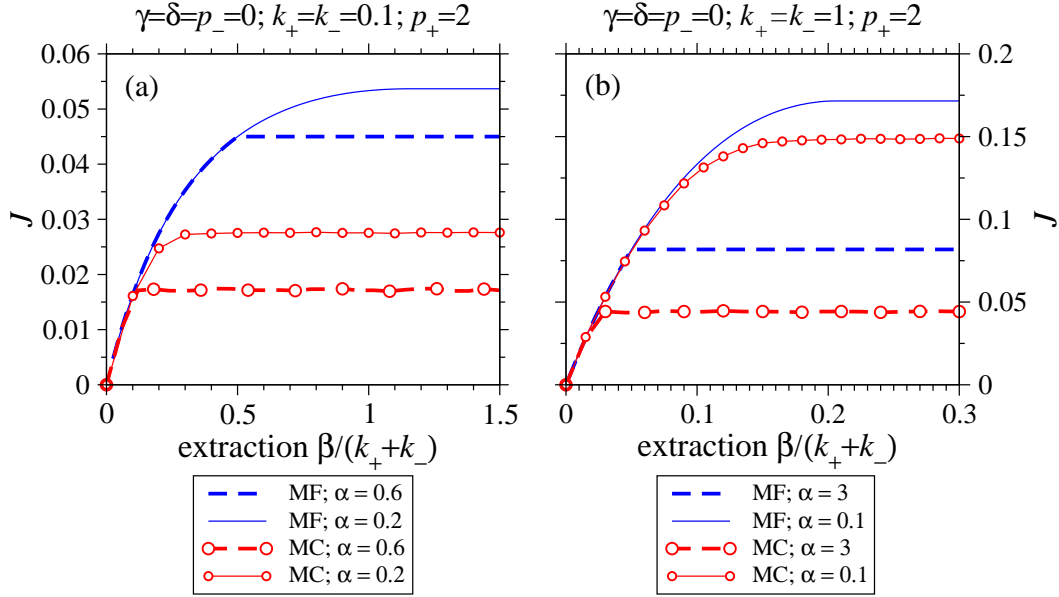}
\caption{Comparison of mean field and $N=50$ Monte-Carlo simulation 
(filled circles) results for the steady-state current
with $\g=\d=p_{-}=0$ and $p_{+}=2$. (a) Currents as a 
function of $\b/(k_{+}+k_{-})$ for $k_{+}=k_{-}=0.1$ and 
$\a = 0.2, 0.6$. (b). Currents for  $k_{+}=k_{-}=1$ and 
$\a=0.1, 3$. Thin solid curves correspond to the smaller 
values of $\a$, while thick dashed curves depict currents 
with the larger $\a$.}
\end{center}
\label{ASYMP0}
\end{figure}

Figures 3a,b show slices (for $N=50$, $k_{\pm}=0.1,1$, and fixed $\a$) through
the phase diagram of the totally asymmetric case (Fig. 2).  The mean field
currents are represented by the thin solid and thick dashed curves for different
values of $\a$ in each plot.  Mean field theory seems to predict a first order
transdition between the $J_{L}^{-}$ and $J_{R}^{-}$ phases and a higher order
transition to the $J_{M}^{-}$ phase. The curves labelled with $\bigcirc$
symbols represent steady-state currents computed from the MC simulations for
the same fixed values of $\a$. The mean field solutions always overestimate the
currents with the error especially bad for small $k_{\pm}$ (compare Fig. 3a
with Fig. 3b). This suggests correlations are suppressed when water flipping is
fast, rendering the mean field approximation more accurate. A similar qualitative
trend arises in the symmetric case.  


Figure 4 shows the steady-state currents for $N=100$ in the special symmetric
case ($p_{+}=p_{-}=1$) as a function of the equal flip rates $k=k_{+}=k_{-}$. 
The asymptotic mean-field currents for ($\a=0.2, \d=0.0$) and ($\a=1.0, \d =
0.0$) are plotted with thin and thick dashed curves, respectively. The mean-field
results again overestimate the steady-state current $J$, and become
increasingly accurate for large flip rates.  
The $k\rightarrow \infty$ limit in which mean field theory is asymptotically exact
gives

\begin{equation}
\fl J \sim {p_{-} \over N}\left[{\a\b-\g\d \over (2\b+\d)(2\g+\a)} - {p_{-}\over k}
{\a^{2}\b(\b+\d)-\d^{2}\g(\g+\a)\over (2\b+\d)^{2}(2\g+\a)^{2}} 
+ O\left({p_{-}^{2} \over k^{2}}\right)\right].
\label{KINF}
\end{equation}  

\begin{figure}
\begin{center}
\includegraphics[height=2.7in]{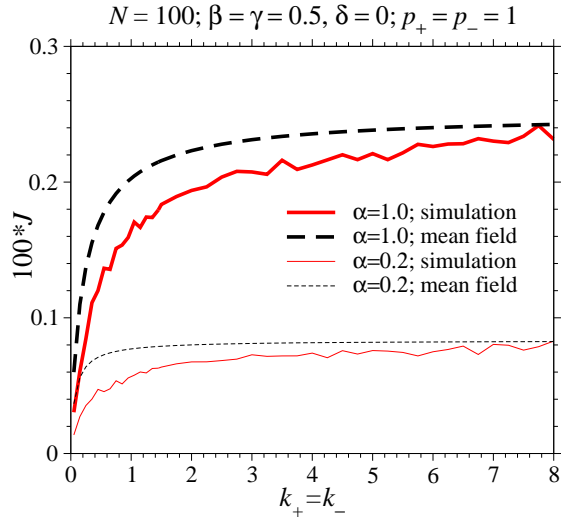}
\caption{Mean field (dashed) and simulated (solid) steady-state currents for 
a ``symmetric'' $(k_{-}p_{+}=k_{+}p_{-})$ chain with $N=100, 
\b = 0.5, \delta=0, p_{+}=p_{-}=1$ and $\a=0.2,1.0$. }
\end{center}
\label{ASYMP}
\end{figure}

Site occupations depicted in Fig. 5a clearly indicate the absence of boundary layers in
the symmetric limit.  The proton concentration is nearly  linear along the chain and is
fairly insensitive to the flipping rate $k$. Only for small $k=0.1$ (thin lines, small
symbols) does the profile deviate slightly from being linear.  Not surprisingly, the mean
field result for the steady-state current is substantially more accurate when $k$ is
large.  Figure 5b shows the correlations

\begin{equation}
S(\s_{i},\s_{i+1}) \equiv \langle (1-\s_{i}^{2})(1-\s_{i+1}^{2})\rangle -\langle 1-\s_{i}^{2}\rangle
\langle 1-\s_{i+1}^{2}\rangle 
\end{equation}

\noindent as functions of the lattice site $i$. We plot $S(+,+)$ and $S(0,+)$ (circles) for
both $k=0.1$ (thin dashed lines)  and $k=10$(thick solid lines). Correlations $S$
containing $\s_{i}= -$ do not vary appreciably when $k$ is changed. Note that $S(+,+)$
for $k=0.1$ is larger than $S(+,+)$ for $k=10$. The larger flipping rates wash out the
correlations and suppresses $S(+,+)$.  Conversely, $S(0,+)$ for small $k=0.1$ is
negative since $0+ \rightarrow -0$ is a relatively fast step that destroys the $0+$
configuration. When $k$ is large, this negative correlation is also supressed.

\begin{figure}
\begin{center}
\includegraphics[height=3.5in]{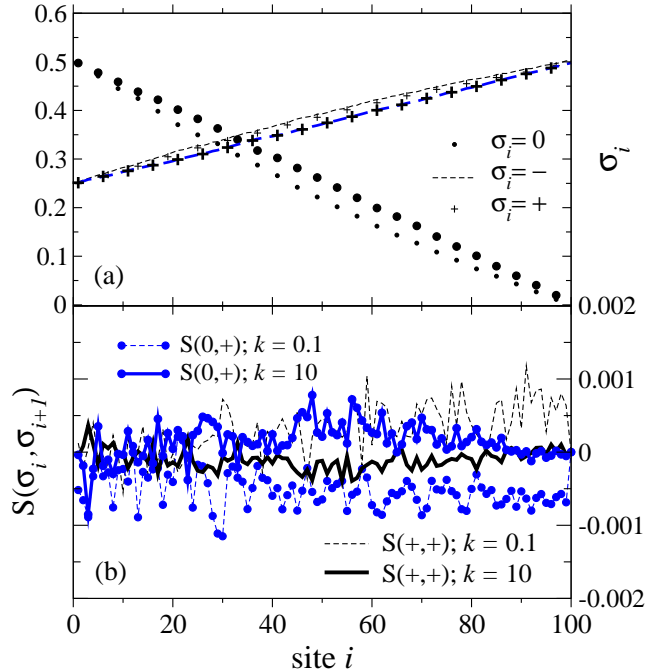}
\caption{(a) Steady-state occupations of $0,-,+$ along the chain for
$\a=1,\b=\g=0.5,\d=0$ and $p_{\pm}=1$. The larger symbols correspond 
to $k_{\pm}=k=10$, while the small, thin symbols indicate 
occupations associated with $k=0.1$. The differences between 
$\s_{i}=0$ and $\s_{i}=-$ are barely discernible.
(b). The correlations $S(+,+)$ and $S(0,+)$ for small($k=0.1$) and large($k=10$)
flip rates.}
\end{center}
\label{CORR}
\end{figure}

Figures 6 contrasts concentration  difference driven currents and potential
difference driven currents. In figure 6a, $100\times J$ is plotted as a function of
$\a$ for $N=20$ and $\delta = 0$.  The current is always sublinear with respect
to $\a$ and increases for small, increasing $k$.  For larger $k$, the current
saturates as the rate limiting step becomes the forward hopping rate $p$.

\begin{figure}
\begin{center}
\includegraphics[height=3.5in]{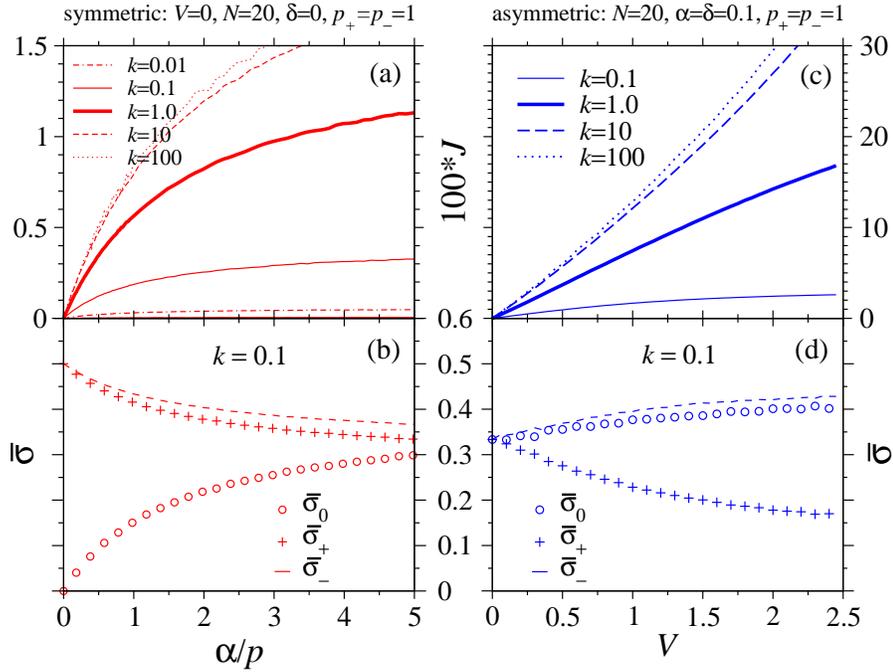}
\caption{(a) $J$ as a function of $\a$ for $N=20$ and $\d=0, p_{\pm}=1$, and 
various $k$. The corresponding chain-averaged populations for $k=0.1$ are shown in (b). 
(c) Currents as a function of driving $V$ defined by (\ref{DRIVING}).
All $J-V$ curves saturate (become sublinear and level off)
at large $V$ when flipping rates $k$ become rate-limiting.
(d) Averaged populations for $k=0.1$ as functions of driving $V$.}
\end{center}
\label{VK}
\end{figure}

The site occupancies, averaged over the entire chain, are plotted in Fig.
6b.  The curves plot $\bar{\sigma}_{0} =
N^{-1}\sum_{i=1}^{N}(1-\s_{i}^{2})$, $\bar{\sigma}_{-} =
(2N)^{-1}\sum_{i=1}^{N}\s_{i}(\s_{i}-1)$, and $\bar{\sigma}_{+} =
(2N)^{-1}\sum_{i=1}^{N}\s_{i}(\s_{i}+1)$, representing the chain-averaged proton, $-$
type, and $+$ type particle densities, respectively.  As the proton entrance rate $\a$ is
increased, the proton concentration in the pore increases at the expense of
both $-$ and $+$ particles. However, since the current to the right increases,
the average population of $+$ particles decreases faster than that of the $-$
particles. For larger $k$, the difference between $-$ and $+$ occupations
vanishes as the high flip rate equalizes their probabilities; the only small
remaining difference arises from the occasional passing of a proton converting
$+$ to $-$.

For potential driven flow, the forward and backward hopping rates were modified
according to 

\begin{equation}
\begin{array}{l}
(\a,\b, p) \rightarrow (\a,\b,p)e^{V/2} \\[13pt]
(\g,\d, q) \rightarrow (\g,\d, q)e^{-V/2}
\label{DRIVING}
\end{array}
\end{equation}

\noindent As a function of $V$, the current $J$ can appear to be sublinear or
superlinear depending on the voltage regime (Fig. 6c).  For small $k$ and sufficiently
large $V$, the rate limiting steps are the $+ \leftrightarrow -$ flips and the current $J$
is sublinear.  The voltage-driven average particle densities for $k=0.1$  (Fig. 6d) also
show increasing proton concentrations and decreasing $+$ particle densities.

\section{Conclusions}

We have formulated an interacting, three-species transport model (epitomized
in Fig.  1) which is motivated by the Grotthuss mechanism \cite{GROTTHUSS}
of proton transfer along water wires. The dynamics resemble those of
two-species ASEPs but with important differences that give rise to very
different behavior. Our main results are summarized by analytic expressions for
the steady-state currents derived within a mean-field approximation.  Currents
in both symmetric (equations \ref{J2} and \ref{J2b}) and asymmetric regimes
(equations \ref{JM}, \ref{JL}, and \ref{JR}) were found, with phases in the totally
asymmetric limit (equation \ref{J0}) analogous to those of the standard ASEP. 
Monte-Carlo simulation agree with mean field analyses only in the limit of large
flipping rates $k_{\pm}$.  Although qualitatively correct, the totally asymmetric
mean field currents described by the phase diagram in Fig. 2, always
overestimates the actual currents, as shown in Fig.  3.  As apparent from Figs. 
3,4, steady-state currents in both symmetric and asymmetric limits are very
sensitive to subtle variations in average occupation and minute differences in
nearest neighbor correlations plotted in Fig. 5.

Our model does not include many important interactions that may be important in
real physical systems.  Examples include dipole-dipole interactions between
adjacent waters, hydronium-hydronium repulsion between two adjacent protons, and
potential induced ordering fields that preferentially orient the water dipoles.
These effects can be readily incorporated by considering next-nearest neighbor
interactions. The more complicated interaction energy can be used in
weighting Monte-Carlo acceptance steps.  

A salient experimental finding in membrane potential-driven water wire proton
conduction is the crossover from sublinear $J-V$ curves to superlinear curves
as the pH in the two equal reservoirs is decreased \cite{EISENMAN80}. Multiple
protons on the water wire have been implicated in the crossover to superlinear
behavior.  Unlike previous theories \cite{SHU}, our model allows for multiple
proton entry into the water wire.  In addition to finding a wider regime of
superlinearity for large $k_{\pm}$, we also found (not shown) stronger
superlinearity when the injection rates ($\a,\d$) were large relative to extraction
rates ($\b,\g$).  These results of the simple lattice model display a number of
different regimes, provide relative estimates for  fundamental tranition rates,
and will serve as a starting point for more complicated models of measured
$J-V$ characteristics \cite{CROSS}.

\vspace{3mm}

\noindent {\bf Acknowledgments} The author thanks G. Lakatos 
and J. Rudnick for helpful comments and suggestions.

%
%

%
%

\end{document}